\documentclass[aps,prl,twocolumn,groupedaddress,superscriptaddress]{revtex4-2}
\usepackage{graphicx}
\usepackage{amssymb,amsmath}
\usepackage{epstopdf}
\usepackage{float}
\usepackage{xcolor}
\usepackage{pifont}
\usepackage{bm}
\usepackage{mathrsfs}
\usepackage{bbold}
\usepackage{upgreek}

\usepackage{times}
\usepackage{url}
\usepackage{xcolor}
\usepackage{gensymb}

\begin{document}

\title{Controllable Production of Degenerate Fermi Gases of $^6$Li Atoms in the 2D-3D Crossover}

\author{Hongwei Gong}
\affiliation{School of Physics and Astronomy, Sun Yat-sen University, Zhuhai, Guangdong, 519082, China}
\author{Haotian Liu}
\affiliation{School of Physics and Astronomy, Sun Yat-sen University, Zhuhai, Guangdong, 519082, China}
\author{Bolong Jiao}
\affiliation{School of Physics and Astronomy, Sun Yat-sen University, Zhuhai, Guangdong, 519082, China}
\author{Haoyi Zhang}
\affiliation{School of Physics and Astronomy, Sun Yat-sen University, Zhuhai, Guangdong, 519082, China}
\author{Qinxuan Peng}
\affiliation{School of Physics and Astronomy, Sun Yat-sen University, Zhuhai, Guangdong, 519082, China}
\author{Shuai Peng}
\affiliation{School of Physics and Astronomy, Sun Yat-sen University, Zhuhai, Guangdong, 519082, China}
\author{Tangqian Shu}
\affiliation{School of Physics and Astronomy, Sun Yat-sen University, Zhuhai, Guangdong, 519082, China}
\author{Hang Yu}
\affiliation{School of Physics and Astronomy, Sun Yat-sen University, Zhuhai, Guangdong, 519082, China}
\author{Yan Zhu}
\affiliation{School of Physics and Astronomy, Sun Yat-sen University, Zhuhai, Guangdong, 519082, China}
\author{Jiaming Li}
\email[]{lijiam29@mail.sysu.edu.cn} \affiliation{School of Physics and Astronomy, Sun Yat-sen
University, Zhuhai, Guangdong, 519082, China} \affiliation{Center of Quantum Information
Technology, Shenzhen Research Institute of Sun Yat-sen University, Shenzhen, Guangdong, China
518087} \affiliation{State Key Laboratory of Optoelectronic Materials and Technologies, Sun Yat-Sen
University, Guangzhou 510275, China}
\author{Le Luo}
\email[]{luole5@mail.sysu.edu.cn} \affiliation{School of Physics and Astronomy, Sun Yat-sen
University, Zhuhai, Guangdong, 519082, China} \affiliation{Center of Quantum Information
Technology, Shenzhen Research Institute of Sun Yat-sen University, Shenzhen, Guangdong, China
518087}
\affiliation{State Key Laboratory of Optoelectronic Materials and Technologies, Sun Yat-Sen
University, Guangzhou 510275, China}

\date{\today}

\begin{abstract}

The many-body physics in the dimensional crossover regime attracts much attention in cold atom
experiments, but yet to explore systematically. One of the technical difficulties existed in the
experiments is the lack of the experimental technique to quantitatively tune the atom occupation
ratio of the different lattice bands. In this letter, we report such techniques in a process of
transferring a 3D Fermi gas into a 1D optical lattice, where the capability of tuning the
occupation of the energy band is realized by varying the trapping potentials of the optical dipole
trap (ODT) and the lattice, respectively. We could tune a Fermi gas with the occupation in the
lowest band from unity to 50$\%$ quantitatively. This provides a route to experimentally study the
dependence of many-body interaction on the dimensionality in a Fermi gas.
\end{abstract}

\maketitle

Over the past decades, experiments of 2D Fermi gases have attracted significant interest since
they provide a highly controllable tool to explore many-body physics in the flatland. The experimental progress includes, but not limited to, the preparation and production of 2D Fermi gas
~\cite{Martiyanov2010, Dyke2011}, the observation of dimension-modified interaction~\cite{Gunter2005} and polaron~\cite{Zhang2012, Koschorreck2012}, thermodynamic measurements of the equation of the states~\cite{Makhalov2014, Martiyanov2016, Boettcher2016, Fenech2016}, radio-frequency spectrum
~\cite{Feld2011, Frohlich2011, Sommer2012, Baur2012, Ong2015}, observation of paring and
Berezinskii-Kosterlitz-Thouless phase transition~\cite{Ries2015, Murthy2015, Mitra2016, Murthy2018},
the measurements of the transport properties~\cite{Koschorreck2013,Bohlen2020,Sobirey2021}, the measurements of the collective mode for quantum anomaly~\cite{Vogt2012, Holten2018, Peppler2018, Murthy2019}, and the realization of the Josephson junction~\cite{Luick2020}. However, most of these experiments were focused on the dependence of many-body physics on the interaction strength as well as the temperature, the rich physics of such dependence on the dimensionality crossover is lesser explored~\cite{Dyke2011, Sommer2012, Cheng2016, Peppler2018}, because of lacking a highly controllable way to produce Fermi gases in the 2D-3D crossover. In contrast, there exist fascinating physics in this crossover regime, such as a contact with non-integer dimensionality, the dimensional evolution of quantum anomaly, and the possible higher temperature superfluidity, etc.

In this letter, we report the production of degenerate Fermi gases of $^6$Li atoms in the 2D-3D
crossover in a controllable way. The scheme applies a transfer of a 3D Fermi gas in an optical
dipole trap (ODT) into a one-dimensional (1D) optical lattice, where the capability of tuning the
occupation of the energy band is realized by varying the atom temperature in the ODT, so that the band occupation can be determined by the relation between the energy distribution of the atom and the energy gap along the tight confinement direction of the lattice.

\begin{figure}[htbp]
    \begin{center}
        \includegraphics[width=\columnwidth, angle=0]{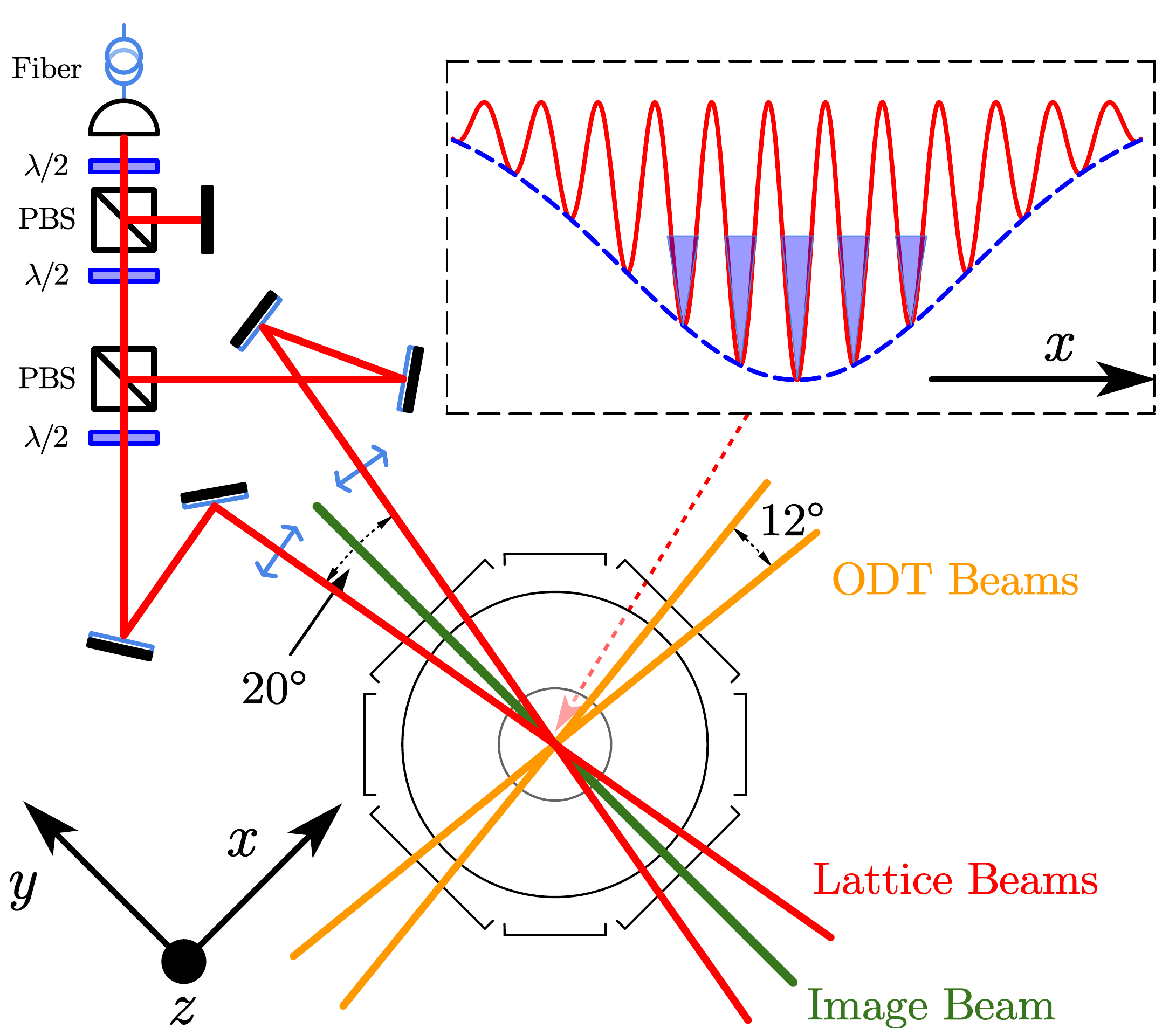}
        \caption{The schematic diagram of the experimental setup for 1D optical lattice, the ODT, and absorption image beam. The lattice potential is shown in the inset. PBS is the polarizing beam splitter. $\lambda /2$ is the half-wave plate.
        }    \label{p:optical_path}
    \end{center}
\end{figure}

The schematic diagram of the experimental setup is shown in Fig.~\ref{p:optical_path}. A single frequency, linearly polarized laser at $\lambda =1064$ $\mathrm{nm}$ with a typical linewidth of 1 $\mathrm{kHz}$(Coherent Prometheus 100NE) is used to form a 1D optical lattice. After passing through an acousto-optic modulator (AOM, CETC SGT80-1064-1TA), the laser is coupled into a single-mode polarization-maintaining fiber with a coreless end cap to ensure that the induced optical lattice has a good Gaussian distribution under a higher laser power. To keep the power of the two beams for the lattice to be stable, two pairs of the half-wave plate and PBS are placed behind the collimator. Then the beam is split into two beams of the same power and polarization ($z$-axis). The two beams travel through roughly the same optical path and are focused by the lens into the center of the experimental chamber. Thus the potential for trapping atoms in the center of the chamber is a combination of an optical lattice in the $x$-axis (lattice axial direction) and Gaussian-shape confinement in the $yz$-plane (lattice radial direction). At the bottom of each lattice site, the potential is nearly harmonic. In our experiment, the angle between the two crossed beams is $2\theta =20\degree$. The power of each beam is 0.53 $\mathrm{W}$, and the Gaussian radius of the beam is about 123 $\mathrm{\upmu m}$ in the center of the chamber. The lattice constant $d={{\lambda}/{\left( 2\sin \theta \right)}}=3.06$ $\mathrm{\upmu m}$. The maximum trap depth value is $127.3E_R$, where $E_R={{\left( 2\pi \hbar \sin \theta \right) ^2}/{\left( 2m\lambda ^2 \right)}}$ is the recoil energy. $\hbar$ is Planck’s constant. $m$ is the mass of the $^6$Li atom. Atoms are typically trapped at 1/3 of the diameter of a Gaussian beam, as shown in Fig.~\ref{p:optical_path}. In our case, the axial length forming the lattice is about 83 $\mathrm{\upmu m}$, forming about 27 lattice sites. The average depth of the lattice is $U_0=118.6E_R=5.4\hbar \omega _x=k_{\mathrm{B}}\cdot 5$ $\mathrm{\upmu K}$, where $k_{\mathrm{B}}$ is Boltzmann’s constant. In the following, we ignored the difference between the lattice sites and treat each lattice site as the same using the average lattice depth. Each lattice site is strongly anisotropic, the ratio of trap frequencies of the central site is $\omega _x:\omega
_y:\omega _z\approx 514:1:5.76$. When the trap depth value is $60E_R$, the tunneling time is 5 $\mathrm{s}$~\cite{Martiyanov2010}, which is much longer than the typical 2D experimental time of 100 $\mathrm{ms}$. Therefore, when the trap depth is more than $60E_R$, tunneling between the different lattice sites is negligible, and the Fermi gas remains kinematically two-dimensional.

\begin{figure}[htbp]
    \begin{center}
        \includegraphics[width=\columnwidth, angle=0]{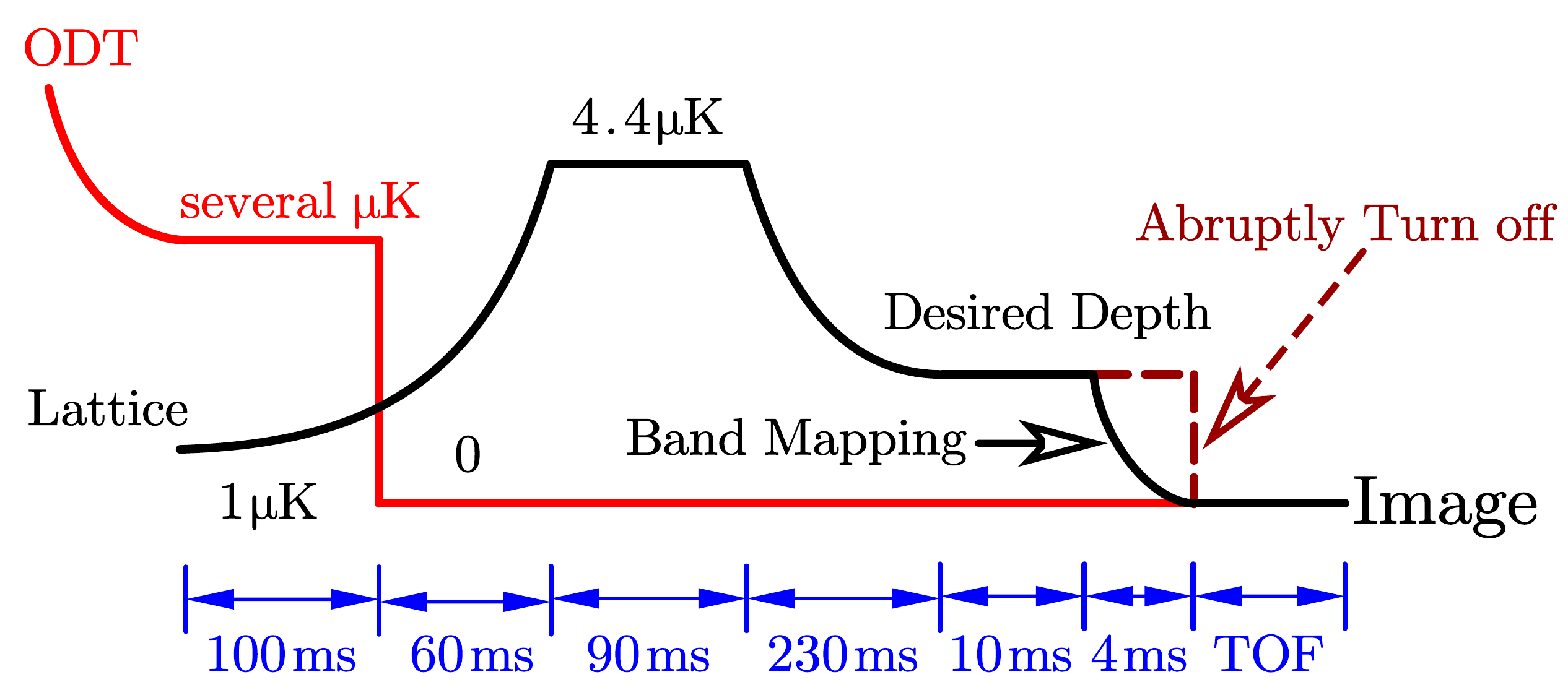}
        \caption{Typical experimental sequence for the production of quasi-2D Fermi gases. We keep the lattice depth for 10 $\mathrm{ms}$ and wait for thermal equilibrium when the lattice depth reaches the desired depth. After that, we choose to abruptly turn off the lattice or do band mapping, then let the atom cloud expand and take the images.
        }    \label{p:typical experimental sequence}
    \end{center}
\end{figure}

We produce the ultracold Fermi gases in the ODT~\cite{Li2016, Li2018} and transfer gases to a 1D optical lattice to produce quasi-2D degenerate Fermi gases. The crossed beams ODT with a crossed angle of $12\degree$ is made by a 100 $\mathrm{W}$ fiber laser at 1064 $\mathrm{nm}$ (IPG Photonics YLR-100-LP). The Gaussian radius of the beams at the center of the chamber is 37 $\mathrm{\upmu m}$. The maximum ODT depth is $k_{\mathrm{B}}\cdot 5.6$ $\mathrm{mK}$. As shown in Fig.~\ref{p:typical experimental sequence}, a typical experimental procedure is described in the following steps. First, about $2\times 10^8$ atoms at a temperature of 300 $\mathrm{\upmu K}$ are trapped in a magneto-optical trap (MOT). Second, we transfer the atoms from MOT to ODT for evaporative cooling, typically lowering the ODT depth to several $\mathrm{\upmu K}$. In our experiment, we use the two lowest-energy hyperfine ground states of $^6$Li, $\left| 2^2S_{1/2},F={{1}/{2}},m_F={{1}/{2}} \right> $ and $\left| 2^2S_{1/2},F={{1}/{2}},m_F={{-1}/{2}} \right> $, usually labeled $\left| 1 \right> $ and $\left| 2 \right> $, respectively. A radio-frequency pulse is then applied to produce a 50:50 mixture of atoms in $\left| 1 \right> $ and $\left| 2 \right> $. At this point, there are about $5\times 10^5$ atoms per spin state in the ODT. Third, starting from the ODT depth dropping to several $\mathrm{\upmu K}$, we ramp up the lattice depth from 1 $\mathrm{\upmu K}$ to 4.4 $\mathrm{\upmu K}$ with an exponential ramp of 160 $\mathrm{ms}$, while the ODT is kept for 100 $\mathrm{ms}$ before turning off. Afterward, the lattice depth is kept stationary for 90 $\mathrm{ms}$. Then the lattice depth is exponentially decreased to the desired depth at 230 $\mathrm{ms}$ for further evaporative cooling. Eventually, we have about $1.5\times10^5$ atoms per spin state in the lattice.

In ultracold Fermi gases, we can realize two different 2D-3D dimensional crossover mechanisms. One is band occupation dimensional crossover and the other is interparticle scattering dimensional crossover. Band occupation dimensional crossover refers to the atoms occupying different energy bands in the lattice. A noninteracting Fermi gas is kinematically two-dimensional when the effective global chemical potential of the gas (compare to the average lattice depth) $\mu<3/2\hbar \omega _x$ and the atoms occupy only the lowest energy band in the lattice, where $\hbar \omega _x$ is the energy level gap from the ground state to the first excited state in the tight-binding direction. In this case, the atom has only two energy level degrees of freedom in the weakly bound direction. At zero temperature, the maximum number of atoms allowed in each lattice site satisfies the 2D condition is $N_\mathrm{2D}=\eta \left( \eta +1 \right) /2$~\cite{Dyke2011}, where $\eta =\omega _x/\sqrt{\omega _y\omega _z}$. If $\mu>3/2\hbar \omega _x$, atoms occupy multiple energy bands and the Fermi gases are in the 2D-3D crossover. Interparticle scattering dimensional crossover means that the characteristic length of the binding potential in the tight-binding direction is larger or smaller than the scattering length between particles. The scattering properties between particles will also be different in the 2D-3D crossover.

In our experiment, although $N_\mathrm{2D}$ is $2.3\times10^4$ larger than the atoms in each lattice site with a typical value of $5.5\times10^3$, we can still quantitatively change the atom occupation in the higher bands. To increase the occupation of the higher bands occupation, we increase the ODT depth when the atoms are transferred from the ODT, so that the atoms have enough energy to occupy the higher bands. On the contrary, to decrease the occupation of the higher bands, we lower the lattice depth after the atoms are transferred to the lattice, to let atoms in the higher bands evaporate.

\begin{figure*}[htbp]
    \begin{center}
        \includegraphics[width=2\columnwidth]{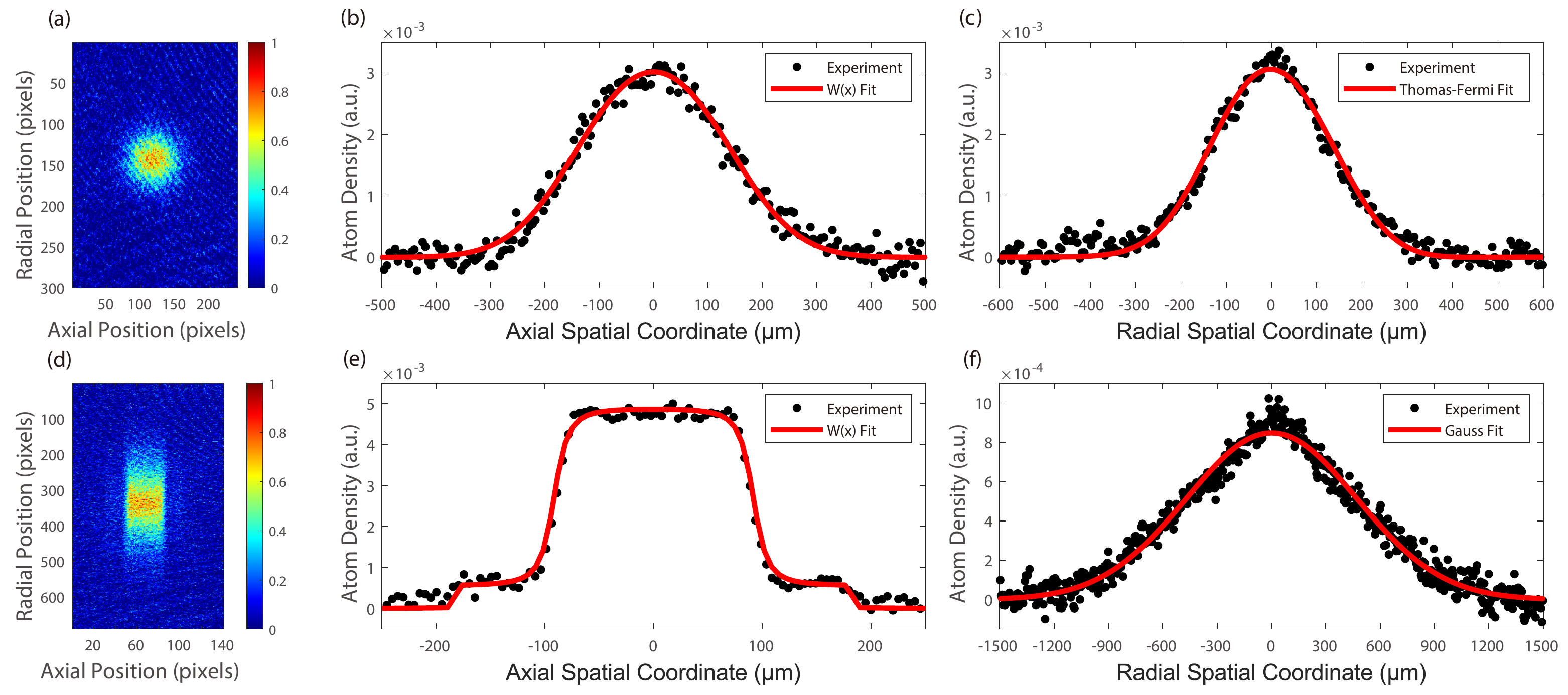}
        \caption{The column density (a)(d), axial (b)(e) and radial (c)(f) atom spatial density distribution (normalized) after abruptly turn-off and band mapping, respectively.
        }  \label{p:ATO_BM_distribution}
    \end{center}
\end{figure*}

We get the band occupation information by taking the absorption images. So far, it is not possible for us to directly take individual images of each lattice site because of the image resolution. We took two kinds of images. One is to abruptly turn off the lattice, let the atom cloud expand for the time of flight (TOF) and then take the images. The other one is band mapping~\cite{Natu2012, Greiner2001, Kohl2005, Miranda2010, Frohlich2011PhD, Cheng2016PhD, Waseem2018}, in which we obtain the information of the chemical potential, and analyze the ratio of the atoms occupying the different energy bands. The band mapping is achieved by gradually ramping down the lattice potential adiabatically over a suitable timescale. This timescale needs to be faster than the tunneling time of the lowest energy band of the lattice, so that the occupations of the different quasi-momentum states remain constant during the lowering process. In addition, this timescale also needs to be slow enough so that different quasi-momentum states can be adiabatically transformed into the corresponding momentum states. By doing this, the Bloch waves in the lattice will adiabatically transform into the plane waves. Correspondingly, the quasi-momentum distribution of atoms in the lattice is transformed into the momentum distribution of the atoms when the lattice is off. Following a ballistic expansion after the band mapping, the momentum distribution is then transformed into the spatial density distribution of the atoms.

\begin{figure*}[htbp]
    \begin{center}
        \includegraphics[width=1.8\columnwidth, angle=0]{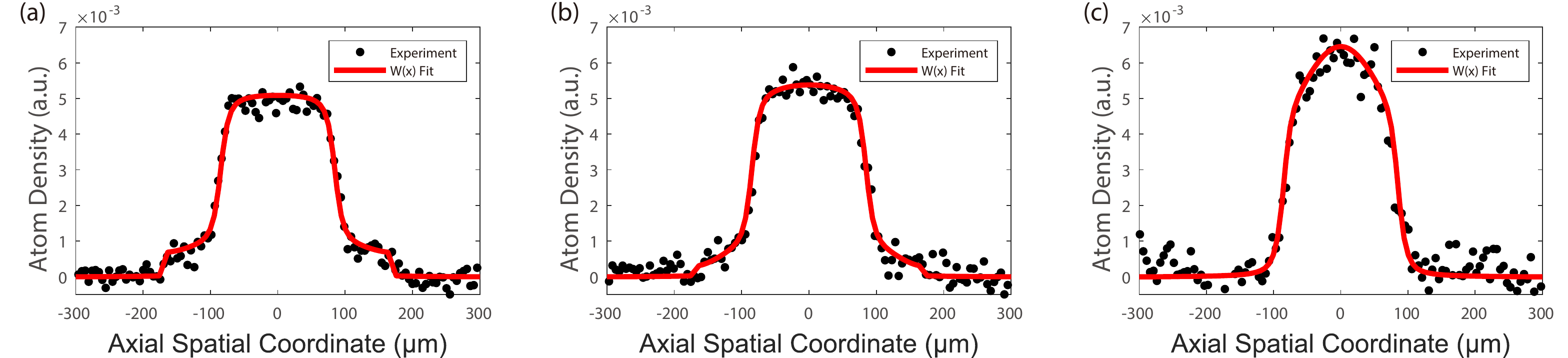}
        \caption{
The atom spatial density distribution along the lattice axial direction when the lattice depth changes. (a) The lattice depth is $14.1E_R$, also similar in Fig.~\ref{p:ATO_BM_distribution}, the fitting give $\mu = 13.7\pm 0.2E_R>3/2\hbar \omega _x=11.3E_R$, $T/T_F=0.70\pm 0.10$, $T=99\pm 14$ $\mathrm{nK}$, there are about $87\%$ of atoms in the lowest band. (b) The lattice depth is $9.6E_R$, $\mu = 10.0\pm 0.1E_R\approx 3/2\hbar \omega _x=9.3E_R$, $T/T_F=0.72\pm 0.03$, $T=75\pm 3$ $\mathrm{nK}$, there are about $91\%$ of atoms in the lowest band. (c) The lattice depth is $4.5E_R$, $\mu = 4.5\pm 0.3E_R<3/2\hbar \omega _x=6.3E_R$, $T/T_F=0.84\pm 0.44$, $T=50\pm 27$ $\mathrm{nK}$, all the atoms are in the lowest band. The numbers of atom are $1.2\times10^5$, $1.0\times10^5$, and $7\times10^4$ from (a) to (c), respectively. As the lattice depth decreases, the atoms in the higher bands escape from the lattice.
        }    \label{p:vary2DU0_BM_distribution}
    \end{center}
\end{figure*}

We first take the image by abruptly turning off the lattice, as shown in Fig.~\ref{p:ATO_BM_distribution}(a). The TOF after abrupt turn off is 5 $\mathrm{ms}$, which is a trade-off to obtain more image data points and maintain a high signal-to-noise ratio. The lattice depth before turn off is $105E_R$, and the bias magnetic field during expansion is 300 $\mathrm{G}$. The axial and radial distribution is Fig.~\ref{p:ATO_BM_distribution}(b) and Fig.~\ref{p:ATO_BM_distribution}(c) respectively. The temperature of the gas is obtained from radial distribution by fitting with 2D Thomas-Fermi distribution. The radial spatial density distribution can be described by~\cite{Martiyanov2010, Kinast2006}
\begin{equation}\label{eq:2DTFdistribution}
n\left( z \right) =-\frac{2N_{a}}{\sqrt{\pi}\sigma _z}\left( \frac{T}{T_F} \right) ^{\frac{3}{2}}\mathrm{Li}_{\frac{3}{2}}\left[ -\exp \left( \frac{{{\mu}/{E_F-{{z^2}/{\sigma _{z}^{2}}}}}}{{{T}/{T_F}}} \right) \right]
\end{equation}
where $\sigma _z=\sqrt{{{2E_F}/{\left( m\omega _{z}^{2} \right)}}}$ is the Fermi radius. $E_F=\hbar \bar{\omega}\sqrt{2N_{a}}$ is the Fermi energy, $\bar{\omega}=\sqrt{\omega _y\omega _z}$. $N_{a}$ is the number of atoms per spin state in a lattice site. $T_F={{E_F}/{k_B}}$ is the Fermi temperature. $\mathrm{Li}\left( x \right) $ is the Polylogarithm. The $T/T_F$ fitted by 2D Thomas-Fermi distribution in Fig.~\ref{p:ATO_BM_distribution}(c) is $0.66\pm 0.06$, which indicates the temperature is $359\pm 31$ $\mathrm{nK}$.

Second, we implement the band mapping by ramping down the lattice potential to $0.01U_0=1.186E_R$ with an exponential ramp of 2.8 $\mathrm{ms}$, as shown in Fig.~\ref{p:ATO_BM_distribution}(d). When the lattice depth is $1.186E_R$, the absolute depth of the lattice is not enough to trap the atoms any more, and the atoms begin to expand. After that, it expands freely before taking the images. The TOF after band mapping is 12.8 $\mathrm{ms}$. The axial and radial distribution is Fig.~\ref{p:ATO_BM_distribution}(e) and Fig.~\ref{p:ATO_BM_distribution}(f), respectively. As shown in Fig.~\ref{p:ATO_BM_distribution}(e), the atoms are step-distribution along the lattice axial direction. For the radial distribution, it's about a Gaussian distribution, as shown in Fig.~\ref{p:ATO_BM_distribution}(f).

We can fit both Fig.~\ref{p:ATO_BM_distribution}(b) and Fig.~\ref{p:ATO_BM_distribution}(e) using the method of Wigner function~\cite{Case2008, Cheng2016PhD}. The fitting equation is $W\left(x\right)$ by replacing the Wigner function $W\left(p\right)$ with $p={m \omega _{mag}x}/{\sin \left( \omega _{mag}t \right)}$. Since the gases expansion is affected by the magnetic field curvature as a harmonic trap, we use the equations of motion for a classical particle in a harmonic trap to describe the gas motion by ignoring the initial position distribution of the atoms, where $\omega _{mag}$ is the frequency of the magnetic trapping potential in the lattice axial direction.

The Wigner function related to the momentum distribution of atoms in the lattices is given by
\begin{equation}\label{eq:momentum distribution}
W\left( p \right) =\frac{1}{\hbar}\sum_{\alpha ,q}{P_{\alpha}^{\mu}\left( q \right) \sum_G{\left| C_{q+G}^{\alpha} \right|^2\delta \left( \frac{p}{\hbar}-q-G \right)}}
\end{equation}
where $\delta $ is the Dirac delta function. $C_{q+G}^{\alpha}$ is a coefficent determined by the Bloch wave function in a 1D optical lattice with
\begin{equation}\label{eq:Bloch wave function}
\psi _{q}^{\alpha}\left( x \right) =\frac{1}{\sqrt{Nd}}\sum_G{C_{q+G}^{\alpha}\exp \left[ i\left( q+G \right) x \right]}
\end{equation}
We obtain the value of $C_{q+G}^{\alpha}$ by solving the time-independent Schrodinger equation in a 1D lattice~\cite{Cheng2016PhD}, where $N$ is the total number of lattice sites. $d$ is the lattice constant. $\alpha $ is the band index. $q={{2n\pi}/{\left( Nd \right)}}$ is the quasi-momentum in the lattice, $-{{\pi}/{d}}\leqslant q\leqslant {{\pi}/{d}}$. $G=2n\pi /d$ is the reciprocal lattice vector, $n$ is an integer.

$P_{\alpha}^{\mu}\left( q \right) $ is the probability that the atom is in the Bloch state with quasi-momentum $q$ and band index $\alpha $ when the chemical potential is $\mu $. In our experiment, we treat the gas with zero temperature, so the probability can be described by
\begin{equation}\label{eq:probability}
P_{\alpha}^{\mu}\left( q \right) =\frac{\left[ \mu -E_{\alpha}\left( q \right) \right] ^2}{\sum_{\alpha ,q}{\left[ \mu -E_{\alpha}\left( q \right) \right] ^2}}\varTheta \left[ \mu -E_{\alpha}\left( q \right) \right]
\end{equation}
where $E_{\alpha}\left( q \right)$ is the energy that the atom is in the Bloch state with quasi-momentum $q$ and band index $\alpha $. $\varTheta $ is the Heaviside step function. For Fig.~\ref{p:ATO_BM_distribution}(b), $P_{\alpha}^{\mu}\left( q \right) $ and $C_{q+G}^{\alpha}$ of Eq.~(\ref{eq:momentum distribution}) are both corresponding to the lattice depth before it is abruptly turned off. For Fig.~\ref{p:ATO_BM_distribution}(e), $P_{\alpha}^{\mu}\left( q \right) $ still corresponds to the lattice depth before the tarp is lowered down, but $C_{q+G}^{\alpha}$ corresponds to the lattice depth after band mapping.

\begin{figure}[htbp]
    \begin{center}
        \includegraphics[width=\columnwidth, angle=0]{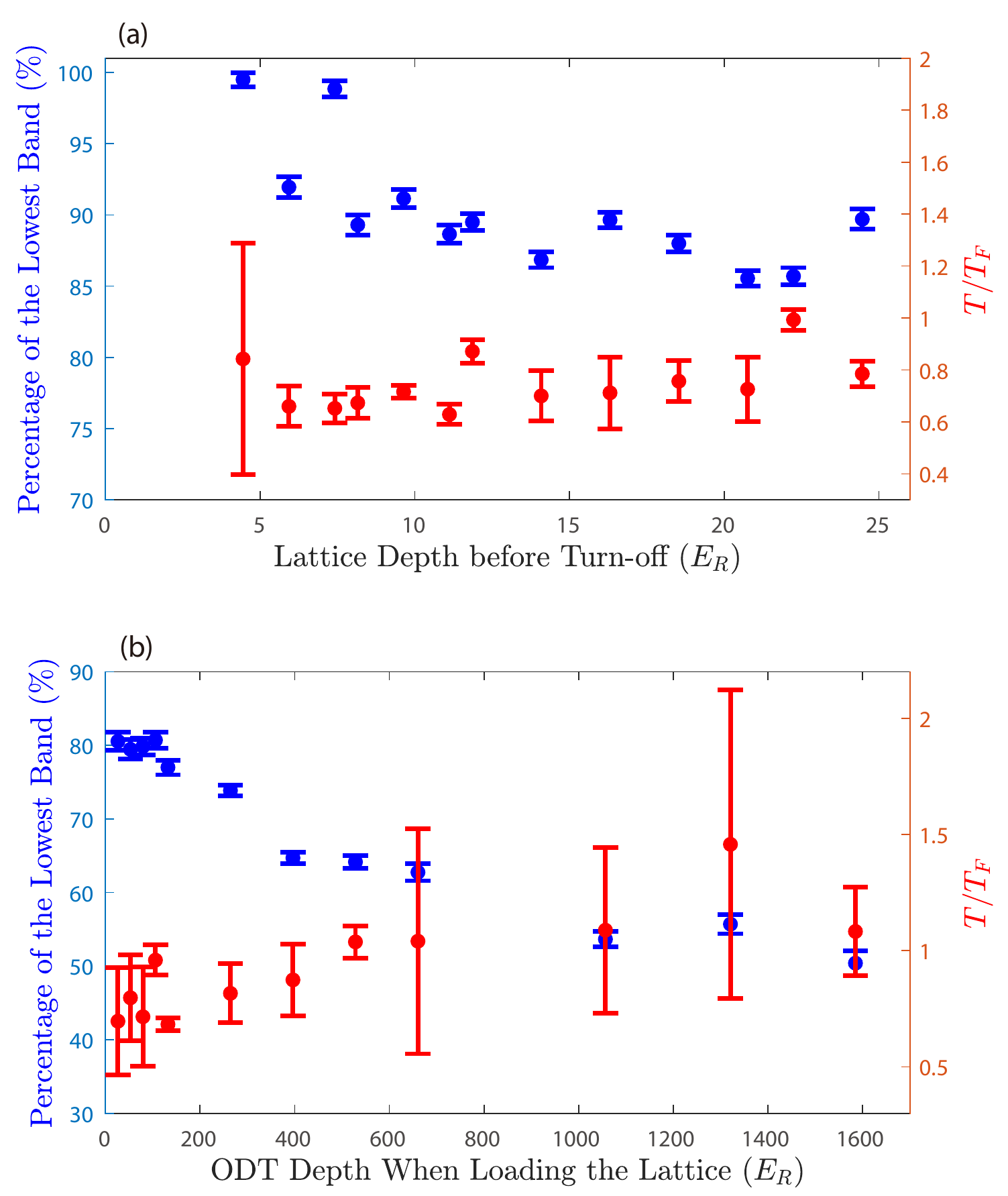}
        \caption{The percentage of atoms occupying the lowest band and $T/T_F$ of Fermi gases versus the lattice depth (a) and the ODT depth when loading the lattice (b). Notice that the percentage data in (b) is derived from the fitting of the axial distribution of abruptly turn-off. Because TOF is less than $T/4$, the fitting result has a systematic error. In this letter, all the errorbar of data is the one standard deviation of image fitting.
        }    \label{p:vary_lowest_band}
    \end{center}
\end{figure}

Fitting Fig.~\ref{p:ATO_BM_distribution}(b) gives $\mu = 44.0\pm 1.0E_R$, there are about $84\%$ of atoms in the lowest band. Fitting Fig.~\ref{p:ATO_BM_distribution}(e) gives $\mu = 39.1\pm 0.3E_R>3/2\hbar \omega _x=30.7E_R$, there are about $90\%$ of atoms in the lowest band. Notice that the TOF of the two images are different, which may result in the different effects of the initial spatial distribution on the momentum distribution. In (e), the TOF is close to $T/4$ corresponding to $\omega _{mag}/{2\pi}=18.8\pm 0.3$ $\mathrm{Hz}$, so that the effect of the initial spatial distribution is minimized~\cite{Cheng2016PhD, Murthy2014}.

To tune the occupation of the energy band, we first vary the lattice depth before it is turned off. As shown in Fig.~\ref{p:vary2DU0_BM_distribution}, the lattice is lower down from (a) to (c), and the atoms occupying the higher bands escape from the trap, which leads to an increase in the percentage of atoms occupying the lowest band. We load a lattice of 25$E_R$ with an ODT around 80$E_R$, where most atoms occupy the lowest two lattice bands, then we decrease the lattice depth to change the percentage of atoms in the lowest band. In Fig.~\ref{p:vary_lowest_band}(a), we plot the dependence of the percentage of atoms occupying the lowest band and $T/T_F$ of Fermi gas on lattice depth. Above 8$E_R$, the proportion of the atom occupancy in the lowest band does not vary significantly with the decrease of the trap depth, Because the lattice depth is well larger than the energy of the excited bands, so the weakly-interacting atoms have no significant evaporative effect, and $T/T_F$ is no significant change. Below 8$E_R$, it is observed that the percentage of atoms in the lowest band increase with the decrease of the lattice depth. The reason is that when the lattice depth approaches 7.5$E_R$, it will not support the first excited band any more, so most atoms in the excited bands evaporate, and almost all the atoms are in the lowest band.

The second way to tune the occupation of the energy band is to vary the gas temperature in the ODT. The gas temperature is varied by changing the ODT depth when loading the lattice, where the higher lattice depth gives a higher gas temperature. For example, we load a lattice of 119$E_R$ with an ODT around 26$E_R$, where we increase the ODT depth to change the percentage of atoms occupying the lowest band. As shown in Fig.~\ref{p:vary_lowest_band}(b), when the ODT depth increase, the gas is hotter and more atoms occupy higher energy bands, resulting in a significant decrease of the percentage of atoms in the lowest band. Since the $T/T_F$ of the gas in the ODT is higher, we observe that the $T/T_F$ in the lattice is also higher.

In summary, we have produced a quasi-2D degenerate Fermi gas by transforming a Fermi gas in the ODT into a 1D optical lattice. We develop two methods to quantitatively control the percentage of atoms in the lowest lattice band either by varying the lattice depth or changing the ODT depth during the lattice loading. By varying the lattice depth, we could prepare a Fermi gas with the occupation in the lowest band from unity to 80$\%$. For varying the ODT depth, we could produce Fermi gases with the occupation in the lowest band from 80$\%$ to 50$\%$. These methods together could quantitatively control the ultracold Fermi gas in the 2D-3D crossover and promote further research on the relationship between many-body interaction and the dimensionality of the system.

\section*{Acknowledgements}
This work is supported by the National Natural Science Foundation of China under Grant No.11774436, No.11804406, and No.12174458, the Key-Area Research and Development Program of Guangdong Province under Grant
No.2019B030330001, the Central-leading-local Scientific and Technological Development Foundation under Grant No.2021Szvup172, and the Fundamental Research Funds for the Central Universities (Sun Yat-sen University, 2021qntd28). Le Luo receives support from Guangdong Province Youth Talent Program under Grant No.2017GC010656.

%


\begin{thebibliography}{38}
    \expandafter\ifx\csname
    natexlab\endcsname\relax\def\natexlab#1{#1}\fi
    \expandafter\ifx\csname bibnamefont\endcsname\relax
    \def\bibnamefont#1{#1}\fi
    \expandafter\ifx\csname bibfnamefont\endcsname\relax
    \def\bibfnamefont#1{#1}\fi
    \expandafter\ifx\csname citenamefont\endcsname\relax
    \def\citenamefont#1{#1}\fi
    \expandafter\ifx\csname url\endcsname\relax
    \def\url#1{\texttt{#1}}\fi
    \expandafter\ifx\csname urlprefix\endcsname\relax\def\urlprefix{URL
    }\fi \providecommand{\bibinfo}[2]{#2}
    \providecommand{\eprint}[2][]{\url{#2}}

    \bibitem{Martiyanov2010}
    \bibinfo{author}{K. Martiyanov},
    \bibinfo{author}{V. Makhalov}, and
    \bibinfo{author}{A. Turlapov},
    \newblock \bibinfo{title} {Observation of a Two-Dimensional Fermi Gas of Atoms},
    \newblock \emph{\bibinfo{journal}{Phys. Rev. Lett.}}
    \textbf{\bibinfo{volume}{105}}, \bibinfo{pages}{030404}
    (\bibinfo{year}{2010}).

    \bibitem{Dyke2011}
    \bibinfo{author}{P. Dyke},
    \bibinfo{author}{E. D. Kuhnle},
    \bibinfo{author}{S. Whitlock},
    \bibinfo{author}{H. Hu},
    \bibinfo{author}{M. Mark},
    \bibinfo{author}{S. Hoinka},
    \bibinfo{author}{M. Lingham},
    \bibinfo{author}{P. Hannaford}, and
    \bibinfo{author}{C. J. Vale},
    \newblock \bibinfo{title} {Crossover from 2D to 3D in a Weakly Interacting Fermi Gas},
    \newblock \emph{\bibinfo{journal}{Phys. Rev. Lett.}}
    \textbf{\bibinfo{volume}{106}}, \bibinfo{pages}{105304}
    (\bibinfo{year}{2011}).

    \bibitem{Gunter2005}
    \bibinfo{author}{K. G\"unter},
    \bibinfo{author}{T. St\"oferle},
    \bibinfo{author}{H. Moritz},
    \bibinfo{author}{M. K\"ohl}, and
    \bibinfo{author}{T. Esslinger},
    \newblock \bibinfo{title} {$p$-Wave Interactions in Low-Dimensional Fermionic Gases},
    \newblock \emph{\bibinfo{journal}{Phys. Rev. Lett.}}
    \textbf{\bibinfo{volume}{95}}, \bibinfo{pages}{230401}
    (\bibinfo{year}{2005}).

     \bibitem{Zhang2012}
    \bibinfo{author}{Y. Zhang},
    \bibinfo{author}{W. Ong},
    \bibinfo{author}{I. Arakelyan}, and
    \bibinfo{author}{J. E. Thomas},
    \newblock \bibinfo{title} {Polaron-to-Polaron Transitions in the Radio-Frequency Spectrum of a Quasi-Two-                    Dimensional Fermi Gas},
    \newblock \emph{\bibinfo{journal}{Phys. Rev. Lett.}}
    \textbf{\bibinfo{volume}{108}}, \bibinfo{pages}{235302}
    (\bibinfo{year}{2012}).

    \bibitem{Koschorreck2012}
    \bibinfo{author}{M. Koschorreck},
    \bibinfo{author}{D. Pertot},
    \bibinfo{author}{E. Vogt},
    \bibinfo{author}{B. Fr\"ohlich},
    \bibinfo{author}{M. Feld}, and
    \bibinfo{author}{M. K\"ohl},
    \newblock \bibinfo{title} {Attractive and repulsive Fermi polarons in two dimensions},
    \newblock \emph{\bibinfo{journal}{Nature}}
    \textbf{\bibinfo{volume}{485}}, \bibinfo{pages}{619-622}
    (\bibinfo{year}{2012}).

    \bibitem{Makhalov2014}
    \bibinfo{author}{V. Makhalov},
    \bibinfo{author}{K. Martiyanov}, and
    \bibinfo{author}{A. Turlapov},
    \newblock \bibinfo{title} {Ground-State Pressure of Quasi-2D Fermi and Bose Gases},
    \newblock \emph{\bibinfo{journal}{Phys. Rev. Lett.}}
    \textbf{\bibinfo{volume}{112}}, \bibinfo{pages}{045301}
    (\bibinfo{year}{2014}).

    \bibitem{Martiyanov2016}
    \bibinfo{author}{K. Martiyanov},
    \bibinfo{author}{T. Barmashova},
    \bibinfo{author}{V. Makhalov}, and
    \bibinfo{author}{A. Turlapov},
    \newblock \bibinfo{title} {Pressure profiles of nonuniform two-dimensional atomic Fermi gases},
    \newblock \emph{\bibinfo{journal}{Phys. Rev. A}}
    \textbf{\bibinfo{volume}{93}}, \bibinfo{pages}{063622}
    (\bibinfo{year}{2016}).

    \bibitem{Boettcher2016}
    \bibinfo{author}{I. Boettcher},
    \bibinfo{author}{L. Bayha},
    \bibinfo{author}{D. Kedar},
    \bibinfo{author}{P. A. Murthy},
    \bibinfo{author}{M. Neidig},
    \bibinfo{author}{M. G. Ries},
    \bibinfo{author}{A. N. Wenz},
    \bibinfo{author}{G. Z\"urn},
    \bibinfo{author}{S. Jochim}, and
    \bibinfo{author}{T. Enss},
    \newblock \bibinfo{title} {Equation of State of Ultracold Fermions in the 2D BEC-BCS Crossover Region},
    \newblock \emph{\bibinfo{journal}{Phys. Rev. Lett.}}
    \textbf{\bibinfo{volume}{116}}, \bibinfo{pages}{045303}
    (\bibinfo{year}{2016}).

    \bibitem{Fenech2016}
    \bibinfo{author}{K. Fenech},
    \bibinfo{author}{P. Dyke},
    \bibinfo{author}{T. Peppler},
    \bibinfo{author}{M. G. Lingham},
    \bibinfo{author}{S. Hoinka},
    \bibinfo{author}{H. Hu}, and
    \bibinfo{author}{C. J. Vale},
    \newblock \bibinfo{title} {Thermodynamics of an Attractive 2D Fermi Gas},
    \newblock \emph{\bibinfo{journal}{Phys. Rev. Lett.}}
    \textbf{\bibinfo{volume}{116}}, \bibinfo{pages}{045302}
    (\bibinfo{year}{2016}).

    \bibitem{Feld2011}
    \bibinfo{author}{M. Feld},
    \bibinfo{author}{B. Fr\"ohlich},
    \bibinfo{author}{E. Vogt},
    \bibinfo{author}{M. Koschorreck}, and
    \bibinfo{author}{M. K\"ohl},
    \newblock \bibinfo{title} {Observation of a pairing pseudogap in a two-dimensional Fermi gas},
    \newblock \emph{\bibinfo{journal}{Nature}}
    \textbf{\bibinfo{volume}{480}}, \bibinfo{pages}{75-78}
    (\bibinfo{year}{2011}).

    \bibitem{Frohlich2011}
    \bibinfo{author}{B. Fr\"ohlich},
    \bibinfo{author}{M. Feld},
    \bibinfo{author}{E. Vogt},
    \bibinfo{author}{M. Koschorreck},
    \bibinfo{author}{W. Zwerger}, and
    \bibinfo{author}{M. K\"ohl},
    \newblock \bibinfo{title} {Radio-Frequency Spectroscopy of a Strongly Interacting Two-Dimensional Fermi         Gas},
    \newblock \emph{\bibinfo{journal}{Phys. Rev. Lett.}}
    \textbf{\bibinfo{volume}{106}}, \bibinfo{pages}{105301}
    (\bibinfo{year}{2011}).

    \bibitem{Sommer2012}
    \bibinfo{author}{A. T. Sommer},
    \bibinfo{author}{L. W. Cheuk},
    \bibinfo{author}{M. J. H. Ku},
    \bibinfo{author}{W. S. Bakr}, and
    \bibinfo{author}{M. W. Zwierlein},
    \newblock \bibinfo{title} {Evolution of Fermion Pairing from Three to Two Dimensions},
    \newblock \emph{\bibinfo{journal}{Phys. Rev. Lett.}}
    \textbf{\bibinfo{volume}{108}}, \bibinfo{pages}{045302}
    (\bibinfo{year}{2012}).

    \bibitem{Baur2012}
    \bibinfo{author}{S. K. Baur},
    \bibinfo{author}{B. Fr\"ohlich},
    \bibinfo{author}{M. Feld},
    \bibinfo{author}{E. Vogt},
    \bibinfo{author}{D. Pertot},
    \bibinfo{author}{M. Koschorreck}, and
    \bibinfo{author}{M. K\"ohl},
    \newblock \bibinfo{title} {Radio-frequency spectra of Feshbach molecules in quasi-two-dimensional                 geometries},
    \newblock \emph{\bibinfo{journal}{Phys. Rev. A}}
    \textbf{\bibinfo{volume}{85}}, \bibinfo{pages}{061604(R)}
    (\bibinfo{year}{2012}).

    \bibitem{Ong2015}
    \bibinfo{author}{W. Ong},
    \bibinfo{author}{C. Cheng},
    \bibinfo{author}{I. Arakelyan}, and
    \bibinfo{author}{J. E. Thomas},
    \newblock \bibinfo{title} {Spin-Imbalanced Quasi-Two-Dimensional Fermi Gases},
    \newblock \emph{\bibinfo{journal}{Phys. Rev. Lett.}}
    \textbf{\bibinfo{volume}{114}}, \bibinfo{pages}{110403}
    (\bibinfo{year}{2015}).

    \bibitem{Ries2015}
    \bibinfo{author}{M. G. Ries},
    \bibinfo{author}{A. N. Wenz},
    \bibinfo{author}{G. Z\"urn},
    \bibinfo{author}{L. Bayha},
    \bibinfo{author}{I. Boettcher},
    \bibinfo{author}{D. Kedar},
    \bibinfo{author}{P. A. Murthy},
    \bibinfo{author}{M. Neidig},
    \bibinfo{author}{T. Lompe}, and
    \bibinfo{author}{S. Jochim},
    \newblock \bibinfo{title} {Observation of Pair Condensation in the Quasi-2D BEC-BCS Crossover},
    \newblock \emph{\bibinfo{journal}{Phys. Rev. Lett.}}
    \textbf{\bibinfo{volume}{114}}, \bibinfo{pages}{230401}
    (\bibinfo{year}{2015}).

    \bibitem{Murthy2015}
    \bibinfo{author}{P. A. Murthy},
    \bibinfo{author}{I. Boettcher},
    \bibinfo{author}{L. Bayha},
    \bibinfo{author}{M. Holzmann},
    \bibinfo{author}{D. Kedar},
    \bibinfo{author}{M. Neidig},
    \bibinfo{author}{M. G. Ries},
    \bibinfo{author}{A. N. Wenz},
    \bibinfo{author}{G. Z\"urn}, and
    \bibinfo{author}{S. Jochim},
    \newblock \bibinfo{title} {Observation of the Berezinskii-Kosterlitz-Thouless Phase Transition in an                   Ultracold Fermi Gas},
    \newblock \emph{\bibinfo{journal}{Phys. Rev. Lett.}}
    \textbf{\bibinfo{volume}{115}}, \bibinfo{pages}{010401}
    (\bibinfo{year}{2015}).

    \bibitem{Mitra2016}
    \bibinfo{author}{D. Mitra},
    \bibinfo{author}{P. T. Brown},
    \bibinfo{author}{P. Schau\ss{}},
    \bibinfo{author}{S. S. Kondov}, and
    \bibinfo{author}{W. S. Bakr},
    \newblock \bibinfo{title} {Phase Separation and Pair Condensation in a Spin-Imbalanced 2D Fermi Gas},
    \newblock \emph{\bibinfo{journal}{Phys. Rev. Lett.}}
    \textbf{\bibinfo{volume}{117}}, \bibinfo{pages}{093601}
    (\bibinfo{year}{2016}).

    \bibitem{Murthy2018}
    \bibinfo{author}{P. A. Murthy},
    \bibinfo{author}{M. Neidig},
    \bibinfo{author}{R. Klemt},
    \bibinfo{author}{L. Bayha},
    \bibinfo{author}{I. Boettcher},
    \bibinfo{author}{T. Enss},
    \bibinfo{author}{M. Holten},
    \bibinfo{author}{G. Z\"urn},
    \bibinfo{author}{P. M. Preiss}, and
    \bibinfo{author}{S. Jochim},
    \newblock \bibinfo{title} {High-temperature pairing in a strongly interacting two-dimensional Fermi gas},
    \newblock \emph{\bibinfo{journal}{Science}}
    \textbf{\bibinfo{volume}{359}}, \bibinfo{pages}{452-455}
    (\bibinfo{year}{2018}).

    \bibitem{Koschorreck2013}
    \bibinfo{author}{M. Koschorreck},
    \bibinfo{author}{D. Pertot},
    \bibinfo{author}{E. Vogt}, and
    \bibinfo{author}{M. K\"ohl},
    \newblock \bibinfo{title} {Universal spin dynamics in two-dimensional Fermi gases},
    \newblock \emph{\bibinfo{journal}{Nature Physics}}
    \textbf{\bibinfo{volume}{9}}, \bibinfo{pages}{405-409}
    (\bibinfo{year}{2013}).

    \bibitem{Bohlen2020}
    \bibinfo{author}{M. Bohlen},
    \bibinfo{author}{L. Sobirey},
    \bibinfo{author}{N. Luick},
    \bibinfo{author}{H. Biss},
    \bibinfo{author}{T. Enss},
    \bibinfo{author}{T. Lompe}, and
    \bibinfo{author}{H. Moritz},
    \newblock \bibinfo{title} {Sound Propagation and Quantum-Limited Damping in a Two-Dimensional Fermi Gas},
    \newblock \emph{\bibinfo{journal}{Phys. Rev. Lett.}}
    \textbf{\bibinfo{volume}{124}}, \bibinfo{pages}{240403}
    (\bibinfo{year}{2020}).

    \bibitem{Sobirey2021}
    \bibinfo{author}{L. Sobirey},
    \bibinfo{author}{N. Luick},
    \bibinfo{author}{M. Bohlen},
    \bibinfo{author}{H. Biss},
    \bibinfo{author}{H. Moritz}, and
    \bibinfo{author}{T. Lompe},
    \newblock \bibinfo{title} {Observation of superfluidity in a strongly correlated two-dimensional Fermi             gas},
    \newblock \emph{\bibinfo{journal}{Science}}
    \textbf{\bibinfo{volume}{372}}, \bibinfo{pages}{844-846}
    (\bibinfo{year}{2021}).

    \bibitem{Vogt2012}
    \bibinfo{author}{E. Vogt},
    \bibinfo{author}{M. Feld},
    \bibinfo{author}{B. Fr\"ohlich},
    \bibinfo{author}{D. Pertot},
    \bibinfo{author}{M. Koschorreck}, and
    \bibinfo{author}{M. K\"ohl},
    \newblock \bibinfo{title} {Scale Invariance and Viscosity of a Two-Dimensional Fermi Gas},
    \newblock \emph{\bibinfo{journal}{Phys. Rev. Lett.}}
    \textbf{\bibinfo{volume}{108}}, \bibinfo{pages}{070404}
    (\bibinfo{year}{2012}).

    \bibitem{Holten2018}
    \bibinfo{author}{M. Holten},
    \bibinfo{author}{L. Bayha},
    \bibinfo{author}{A. C. Klein},
    \bibinfo{author}{P. A. Murthy},
    \bibinfo{author}{P. M. Preiss}, and
    \bibinfo{author}{S. Jochim},
    \newblock \bibinfo{title} {Anomalous Breaking of Scale Invariance in a Two-Dimensional Fermi Gas},
    \newblock \emph{\bibinfo{journal}{Phys. Rev. Lett.}}
    \textbf{\bibinfo{volume}{121}}, \bibinfo{pages}{120401}
    (\bibinfo{year}{2018}).

    \bibitem{Peppler2018}
    \bibinfo{author}{T. Peppler},
    \bibinfo{author}{P. Dyke},
    \bibinfo{author}{M. Zamorano},
    \bibinfo{author}{I. Herrera},
    \bibinfo{author}{S. Hoinka}, and
    \bibinfo{author}{C. J. Vale},
    \newblock \bibinfo{title} {Quantum Anomaly and 2D-3D Crossover in Strongly Interacting Fermi Gases},
    \newblock \emph{\bibinfo{journal}{Phys. Rev. Lett.}}
    \textbf{\bibinfo{volume}{121}}, \bibinfo{pages}{120402}
    (\bibinfo{year}{2018}).

    \bibitem{Murthy2019}
    \bibinfo{author}{P. A. Murthy},
    \bibinfo{author}{N. Defenu},
    \bibinfo{author}{L. Bayha},
    \bibinfo{author}{M. Holten},
    \bibinfo{author}{P. M. Preiss},
    \bibinfo{author}{T. Enss}, and
    \bibinfo{author}{S. Jochim},
    \newblock \bibinfo{title} {Quantum scale anomaly and spatial coherence in a 2D Fermi superfluid},
    \newblock \emph{\bibinfo{journal}{Science}}
    \textbf{\bibinfo{volume}{365}}, \bibinfo{pages}{268-272}
    (\bibinfo{year}{2019}).

    \bibitem{Luick2020}
    \bibinfo{author}{N. Luick},
    \bibinfo{author}{L. Sobirey},
    \bibinfo{author}{M. Bohlen},
    \bibinfo{author}{V. P. Singh},
    \bibinfo{author}{L. Mathey},
    \bibinfo{author}{T. Lompe}, and
    \bibinfo{author}{H. Moritz},
    \newblock \bibinfo{title} {An ideal Josephson junction in an ultracold two-dimensional Fermi gas},
    \newblock \emph{\bibinfo{journal}{Science}}
    \textbf{\bibinfo{volume}{369}}, \bibinfo{pages}{89-91}
    (\bibinfo{year}{2020}).

    \bibitem{Cheng2016}
    \bibinfo{author}{C. Cheng},
    \bibinfo{author}{J. Kangara},
    \bibinfo{author}{I. Arakelyan}, and
    \bibinfo{author}{J. E. Thomas},
    \newblock \bibinfo{title} {Fermi gases in the two-dimensional to quasi-two-dimensional crossover},
    \newblock \emph{\bibinfo{journal}{Phys. Rev. A}}
    \textbf{\bibinfo{volume}{94}}, \bibinfo{pages}{031606(R)}
    (\bibinfo{year}{2016}).

    \bibitem{Li2016}
    \bibinfo{author}{J. Li},
    \bibinfo{author}{J. Liu},
    \bibinfo{author}{W. Xu},
    \bibinfo{author}{L. de Melo}, and
    \bibinfo{author}{L. Luo},
    \newblock \bibinfo{title} {Parametric cooling of a degenerate Fermi gas in an optical trap},
    \newblock \emph{\bibinfo{journal}{Phys. Rev. A}}
    \textbf{\bibinfo{volume}{93}}, \bibinfo{pages}{041401(R)}
    (\bibinfo{year}{2016}).

     \bibitem{Li2018}
    \bibinfo{author}{J. Li},
    \bibinfo{author}{J. Liu},
    \bibinfo{author}{L. Luo}, and
    \bibinfo{author}{B. Gao},
    \newblock \bibinfo{title} {Three-Body Recombination near a Narrow Feshbach Resonance in $^{6}\mathrm{Li}        $},
    \newblock \emph{\bibinfo{journal}{Phys. Rev. Lett.}}
    \textbf{\bibinfo{volume}{120}}, \bibinfo{pages}{193402}
    (\bibinfo{year}{2018}).

    \bibitem{Natu2012}
    \bibinfo{author}{S. S. Natu},
    \bibinfo{author}{D. C. McKay},
    \bibinfo{author}{B. DeMarco}, and
    \bibinfo{author}{E. J. Mueller},
    \newblock \bibinfo{title} {Evolution of condensate fraction during rapid lattice ramps},
    \newblock \emph{\bibinfo{journal}{Phys. Rev. A}}
    \textbf{\bibinfo{volume}{85}}, \bibinfo{pages}{061601(R)}
    (\bibinfo{year}{2012}).

    \bibitem{Greiner2001}
    \bibinfo{author}{M. Greiner},
    \bibinfo{author}{I. Bloch},
    \bibinfo{author}{O. Mandel},
    \bibinfo{author}{T. W. H\"ansch},
    \bibinfo{author}{T. Esslinger},
    \newblock \bibinfo{title} {Exploring Phase Coherence in a 2D Lattice of Bose-Einstein Condensates},
    \newblock \emph{\bibinfo{journal}{Phys. Rev. Lett.}}
    \textbf{\bibinfo{volume}{87}}, \bibinfo{pages}{160405}
    (\bibinfo{year}{2001}).

    \bibitem{Kohl2005}
    \bibinfo{author}{M. K\"ohl},
    \bibinfo{author}{H. Moritz},
    \bibinfo{author}{T. St\"oferle},
    \bibinfo{author}{K. G\"unter}, and
    \bibinfo{author}{T. Esslinger},
    \newblock \bibinfo{title} {Fermionic Atoms in a Three Dimensional Optical Lattice: Observing Fermi               Surfaces, Dynamics, and Interactions},
    \newblock \emph{\bibinfo{journal}{Phys. Rev. Lett.}}
    \textbf{\bibinfo{volume}{94}}, \bibinfo{pages}{080403}
    (\bibinfo{year}{2005}).

    \bibitem{Miranda2010}
    \bibinfo{author}{M. de Miranda, Ph.D},
    \newblock \bibinfo{title} {Control of dipolar collisions in the quantum regime},
    (\bibinfo{year}{2010}).

    \bibitem{Frohlich2011PhD}
    \bibinfo{author}{B. Fr\"ohlich, Ph.D},
    \newblock \bibinfo{title} {A Strongly Interacting Two-Dimensional Fermi Gas},
    (\bibinfo{year}{2011}).

    \bibitem{Cheng2016PhD}
    \bibinfo{author}{C. Cheng, Ph.D},
    \newblock \bibinfo{title} {Ultracold Fermi Gases in a Bichromatic Optical Superlattice},
    (\bibinfo{year}{2016}).

    \bibitem{Waseem2018}
    \bibinfo{author}{M. Waseem, Ph.D},
    \newblock \bibinfo{title} {Collisional properties of Fermi gases with p-wave interactions},
    (\bibinfo{year}{2018}).

    \bibitem{Case2008}
    \bibinfo{author}{W. Case},
    \newblock \bibinfo{title} {Wigner functions and Weyl transforms for pedestrians},
    \newblock \emph{\bibinfo{journal}{American Journal of Physics}}
    \textbf{\bibinfo{volume}{76}}, \bibinfo{pages}{937}
    (\bibinfo{year}{2008}).

    \bibitem{Kinast2006}
    \bibinfo{author}{J. Kinast, Ph.D},
    \newblock \bibinfo{title} {Thermodynamics and superfluidity of a strongly interacting fermi gas},
    (\bibinfo{year}{2006}).

    \bibitem{Murthy2014}
    \bibinfo{author}{P. A. Murthy},
    \bibinfo{author}{D. Kedar},
    \bibinfo{author}{T. Lompe},
    \bibinfo{author}{M. Neidig},
    \bibinfo{author}{M. G. Ries},
    \bibinfo{author}{A. N. Wenz},
    \bibinfo{author}{G. Z\"urn}, and
    \bibinfo{author}{S. Jochim},
    \newblock \bibinfo{title} {Matter-wave Fourier optics with a strongly interacting two-dimensional Fermi          gas},
    \newblock \emph{\bibinfo{journal}{Phys. Rev. A}}
    \textbf{\bibinfo{volume}{90}}, \bibinfo{pages}{043611}
    (\bibinfo{year}{2014}).

\end{thebibliography}
\end{document}